\documentclass[10pt,twoside]{aipproc} 
\layoutstyle{6x9}
\begin{document}
\title{Relativistic Bohmian interpretation of quantum mechanics}
\classification{03.65.Ta, 03.65.Pm}
\keywords{Bohmian interpretation, Relativistic quantum mechanics}
\author{Hrvoje Nikoli\'c}{
  address={Theoretical Physics Division, Rudjer Bo\v{s}kovi\'{c} Institute,
           P.O.B. 180, HR-10002 Zagreb, Croatia},
  email={hrvoje@thphys.irb.hr}
}
\begin{abstract}
I present a relativistic covariant version of the Bohmian interpretation of
quantum mechanics and discuss the corresponding measurable predictions.
The covariance is incoded in the fact that the nonlocal quantum potential 
transforms as a scalar, which is a consequence of the fact 
that the nonlocal wave function transforms as a scalar. The  
measurable predictions that can be obtained with the deterministic 
Bohmian interpretation cannot be obtained with the conventional 
interpretation simply because the conventional probabilistic 
interpretation does not work in the case of relativistic 
quantum mechanics.
\end{abstract}

\maketitle

\section{Introduction}

The Bohmian interpretation of quantum mechanics and quantum field theory 
\cite{bohm,bohmPR1,bohmPR2,holPR,holbook} 
is a promising approach towards the solution of the 
problem of measurement in quantum theory. However, two objections 
on this interpretation are often posed. First, it is a {\em nonlocal}
hidden variable theory, so it seems to be in contradiction with
special theory of relativity.
Second, it does not seem to lead to new measurable predictions, so 
its value seems to be more philosophical than scientific. 
Here I review some of my recent results originally presented 
in \cite{nikolfpl1,nikolfpl3} which show 
that the Bohmian interpretation of quantum mechanics 
can be formulated such that 
it is both nonlocal and compatible with special theory of relativity.
Moreover, it turns out that such a formulation leads to 
new measurable predictions, allowing for an experimental 
verification of the relativistic Bohmian interpretation of 
quantum mechanics. Before discussing the relativistic case, 
I also present a short review of the nonrelativistic Bohmian interpretation. 

\section{Nonrelativistic Bohmian interpretation}

Quantum mechanics (QM) is described by the Schr\"odinger equation
\begin{equation}
\left[ \frac{-\hbar^2\nabla^2}{2m}+V \right] \psi=i\,\hbar\partial_t \psi .
\end{equation}
By writing the wave function in the polar form
\begin{equation} 
\psi({\bf x},t)=R({\bf x},t)e^{iS({\bf x},t)/\hbar},
\end{equation}
the complex Schr\"odinger can be written as a set of two 
real equations. One of them is the
quantum Hamilton-Jacobi equation
\begin{equation}
\frac{(\nabla S)^2}{2m}+V+Q=-\partial_tS ,
\end{equation}
while the other one is the conservation equation
\begin{equation}\label{cons}
\partial_t R^2 + \nabla\left( R^2\frac{\nabla S}{m} \right) =0 .
\end{equation}
Here the quantum potential $Q$ is defined as
\begin{equation}
Q\equiv -\frac{\hbar^2}{2m}\frac{\nabla^2 R}{R} .
\end{equation}
The conservation equation implies that $|\psi|^2$ 
can be interpreted as a probability density. 

The quantum Hamilton-Jacobi equation given above takes the same form as the 
classical Hamilton-Jacobi equation, except for the fact that 
the quantum Hamilton-Jacobi equation contains an additional $Q$-term.
This suggests the Bohmian interpretation, according to which 
the particle has a deterministic trajectory ${\bf x}(t)$ that 
satisfies
\begin{equation}\label{Ebohm}
\frac{d {\bf x}}{dt}=\frac{\nabla S}{m}.
\end{equation}
Eq.~(\ref{Ebohm}) is identical to an analogous equation 
in the classical Hamilton-Jacobi theory. Moreover, 
Eq.~(\ref{Ebohm}) combined with the quantum Hamilton-Jacobi equation
implies the quantum Newton equation
\begin{equation}
m\frac{d^2 {\bf x}}{dt^2}=-\nabla(V+Q),
\end{equation}
which takes the same form as the classical Newton equation, 
except for an additional quantum force generated by the 
quantum potential.  

Now consider a statistical ensemble of particle positions, with the 
probability density $\rho({\bf x},t)$. 
The probability density $\rho$ 
satisfies the conservation equation (\ref{cons}) 
with $R^2\rightarrow\rho$. 
If $\rho({\bf x},t_0)=R^2({\bf x},t_0)$ for some initial time $t_0$, 
then (\ref{Ebohm}) together with the conservation equation provides that
$\rho({\bf x},t)=R^2({\bf x},t)$ for {\em all} $t$. 
This provides a statistical but deterministic explanation of the 
rule that $|\psi|^2$ represents the probability density. 
According to the Bohmian interpretation, all  
QM uncertainties
emerge from the ignorance of the actual initial particle position
${\bf x}(t_0)$. In this interpretation, 
there is no need for a wave-function ``collapse". 
 
But why $\rho({\bf x},t_0)=R^2({\bf x},t_0)$ at the initial time $t_0$?
This is because such a distribution  
corresponds to the statistical {\em equilibrium}. 
There are two variants of this claim.  
Valentini explains this through a quantum H-theorem \cite{val}, 
which says that the system will eventually approach the equilibrium 
distribution even if the initial distribution was not the 
equilibrium one.   
D\"urr, Goldstein, and Zangh\`i \cite{durr1,durr2} explain this by invoking 
a typicality argument. 

From the above, it is clear that the
Bohmian interpretation explains the results of measurements 
on probabilities for particle {\em positions}, without any 
theory on quantum mesurements. 
But what about measurements of other variables, such as 
momentum or energy? 
For {\em other} variables, the agreement with the conventional QM rules 
can be obtained {\em only} by considering the theory of quantum mesurements.
To see this, 
assume that we measure a hermitian operator $\hat{A}$. The eigenstates 
$\psi_a({\bf x})$ of this operator satisfy
\begin{equation}
\hat{A}\psi_a({\bf x})=a\psi_a({\bf x}) ,
\end{equation}
where $a$ are the eigenvalues.
The quantum state can be expanded in terms of these eigenstates as
\begin{equation}
\psi({\bf x},t)=\sum_a c_a(t)\psi_a({\bf x}) .
\end{equation}
Now let
${\bf y}$ be the position corresponding to the measuring apparatus.
According to the von Neumann measurement scheme, the interaction 
between the measured system and the measuring apparatus causes 
the entanglement of the form
\begin{equation}
\Psi({\bf x},{\bf y},t)=\sum_a c_a(t)\psi_a({\bf x})\chi_a({\bf y}) .
\end{equation}
In the Bohmian context, the crucial property
of the measuring apparatus is that different $\chi_a({\bf y})$ 
do not overlap:
\begin{equation}
\chi_a({\bf y})\chi_{a'}({\bf y})=0 \;\;{\rm for}\;\; a\neq a' .
\end{equation}
This implies that
\begin{equation}
|\Psi({\bf x},{\bf y},t)|^2=\sum_a |c_a(t)|^2 |\psi_a({\bf x})|^2 
|\chi_a({\bf y})|^2 ,
\end{equation}
where the mixed terms vanish.
By averaging over ${\bf x}$, we obtain
\begin{equation}\label{yt}
\rho({\bf y},t)=\sum_a |c_a(t)|^2 |\chi_a({\bf y})|^2=|\chi({\bf y},t)|^2 ,
\end{equation} 
where
\begin{equation}
\chi({\bf y},t)\equiv \sum_a c_a(t) \chi_a({\bf y}) .
\end{equation}
Now we apply the Bohmian interpretation to both ${\bf x}$ and ${\bf y}$.
The functions $\chi_a({\bf y})$ form nonoverlapping
localized channels, so that a
particle with the position ${\bf x}(t)$ 
enters only one of the localized channels.
From (\ref{yt}), one infers that the probability
for ${\bf y}$ to take a value from the support of $\chi_a({\bf y})$ is
equal to $|c_a(t)|^2$. 
In other words, the probability to {\em measure} the eigenvalue $a$ of the 
operator $\hat{A}$ is $|c_a(t)|^2$. 
We emphasize that even the position is measured in this way. 
Thus the theory of measurement presented above explains the
effective wave-function collapse: 
The wave function remains a superposition, but the particle moves 
in the same way as if $\psi({\bf x})$ collapsed to $\psi_a({\bf x})$. 

The generalization of the results above 
to the many-particle case is straightforward. 
From a many-particle wave function  
$\psi({\bf x}_1, \ldots, {\bf x}_n,t)$ one calculates
the corresponding many-particle quantum potential 
$Q({\bf x}_1, \ldots, {\bf x}_n,t)$.
Consequently, the quantum force on a particle with the trajectory 
${\bf x}_a(t)$, $a=1,\ldots,n$, is equal to
\begin{equation}
{\bf F}_a({\bf x}_1, \ldots, {\bf x}_n,t)=
-\nabla_a Q({\bf x}_1, \ldots, {\bf x}_n,t) .
\end{equation}
The force on one particle depends on the {\em instantaneous} 
position of all other particles. This is a manifestation
of the nonlocality of QM. In this way, the 
Bohmian interpretation is (the simplest!) {\em nonlocal} 
hidden variable interpretation of QM, fully consistent with the Bell theorem! 

At the end of this section, let us shortly discuss the
advantages and disadvantages of the Bohmian interpretation.
The main advantage is the fact that the Bohmian interpretation 
is conceptually the 
most similar to classical mechanics, which makes it conceptually 
very clear and appealing. 
The main disadvantages (emphasized by those who are skeptical 
about the Bohmian intepretation) are the following ones:
\begin{enumerate}
\item The Bohmian intepretation is 
technically more complicated than the conventional interpretation. 
\item This interpretation does not lead to new measurable 
predictions.\footnote{Since the time-observable 
is not well defined in the conventional 
interpretation of QM, there are some indications 
that the Bohmian interpretation may lead to new measurable
predictions on the time-variable. However, since the theory 
of measurements has not been used in the existing attempts to give the 
Bohmian predictions on the time-variable, 
the correcteness of these attempts is dubious.} 
\item The Bohmian hidden variables are nonlocal 
(owing to the instantaneous action at a distance), 
which possibly violates special relativity.
\end{enumerate}  
In the rest of this paper, I will demonstrate that the two 
(the second one and the third one) 
of these three problems can be solved.
  
\section{Relativistic Bohmian interpretation}

A relativistic quantum spinless particle satisfies
the Klein-Gordon equation 
\begin{equation}
(\partial^{\mu}\partial_{\mu}+m^2)\psi(x)=0 ,
\end{equation}
where $x=(t,\bf{x})$, the signature of the metric 
is $(+,-,-,-)$, and we take $\hbar=c=1$.
The quantity $|\psi|^2$ is not conserved, so it cannot be 
interpreted as a probability density. 
The conserved current is 
\begin{equation}
j_{\mu}=i\psi^* \!\stackrel{\leftrightarrow\;}{\partial_{\mu}}\! \psi .
\end{equation}
However, {\em even for positive-frequency solutions}, it is possible that 
$j_0(x)<0$ at some regions of spacetime. 
Consequently, $j_0$ cannot be interpreted as a probability density either. 

The standard resolution of this problem is 
second quantization (known also under the name {\em quantum field theory}), 
where $\psi$ is not a 
wave function describing probabilities, but an observable, represented 
by a field operator $\hat{\psi}$. 
However, there is a problem with this interpretation.
If, at the fundamental level,
$\psi$ should not be interpreted as a wave function
that determines probabilities of particle positions, then
it is not clear
why the probabilistic interpretation of $\psi$ is in 
agreement with experiments for nonrelativistic particles. 

To solve this problem, I propose that both fields and particle positions 
are fundamental entities \cite{nikolfpl1,nikolfpl2}. 
The field operator $\hat{\phi}$ satisfies
\begin{equation}
(\partial^{\mu}\partial_{\mu}+m^2)\hat{\phi}(x)=0 .
\end{equation}
Here
$\hat{\phi}$ is a hermitian (uncharged) field, which provides that
negative densities are not related to a negative charge. 
An $n$-particle wave function is
\begin{equation}\label{npsi}
\psi(x_1,\ldots ,x_n)=(n!)^{-1/2}S_{\{ x_a\} }
\langle 0|\hat{\phi}(x_1)\cdots\hat{\phi}(x_n)|n\rangle ,
\end{equation}
where $|n\rangle$ is a Lorentz-invariant $n$-particle state, 
$|0\rangle$ is the vacuum, and $S_{\{ x_a\} }$ denotes the symmetrization
needed because the field operators do not commute for nonequal times.
Eq.~(\ref{npsi}) relates second quantization with first quantization, 
because the right-hand side contains quantities that refer to 
second quantization (quantum field theory), while the left-hand side 
represents a quantity related to first quantization.
Note that the wave function depends not only on $n$ 
space positions, but also on $n$ times. In this way, 
space and time are treated on an equal footing, which provides the 
relativistic covariance.

One can also introduce $n$ particle currents (one for each $a$) as
\begin{equation}
j^{\mu}_a=i\psi^* \!\stackrel{\leftrightarrow\;}{\partial^{\mu}_a}\! \psi .
\end{equation}
(For other physical aspects of particle currents and their difference 
with respect to charge currents, see \cite{nikolplb,nikolijmpd,nikolgrg}.)
The currents are conserved, i.e., 
\begin{equation}
\partial^{\mu}_a j_{a\mu}=0 .
\end{equation}
The wave function (\ref{npsi}) satisfies the
$n$-particle Klein-Gordon equation
\begin{equation}
\left( \sum_a\partial_a^{\mu}\partial_{a\mu}+nm^2 \right) 
\psi(x_1,\ldots ,x_n)=0 .
\end{equation}
By writing
\begin{equation}
\psi=Re^{iS} ,
\end{equation}
one obtains the
relativistic conservation equation
\begin{equation}
\sum_a\partial_a^{\mu}(R^2\partial_{a\mu}S)=0,
\end{equation}
and the
relativistic quantum Hamilton-Jacobi equation
\begin{equation}
-\frac{\sum_a(\partial_a^{\mu}S)(\partial_{a\mu}S)}{2m} +\frac{nm}{2} +Q=0 ,
\end{equation}
where the
relativistic quantum potential is
\begin{equation}
Q=\frac{1}{2m}\frac{\sum_a\partial_a^{\mu}\partial_{a\mu}R}{R} .
\end{equation}
It is crucial to note that the quantum potential
$Q(x_1,\ldots ,x_n)$ is nonlocal, but relativistic invariant! 

Now we introduce the Bohmian interpretation. We postulate
the Bohmian equation of motion 
\begin{equation}
\frac{dx_a^{\mu}}{ds} = -\frac{1}{m}\partial_a^{\mu}S . 
\end{equation}
This is equivalent to
\begin{equation}\label{bohmj}
\frac{dx_a^{\mu}}{ds} = \frac{j_a^{\mu}}{2m\psi^*\psi} .
\end{equation}
Consequently, the
relativistic quantum Newton equation reads
\begin{equation}
m\frac{d^2x_a^{\mu}}{ds^2}=\partial_a^{\mu}Q .
\end{equation}
The Bohmian equation of motion is covariant, which is
incoded in the fact that
$s$ is not a specific time coordinate, but an auxiliary 
scalar parameter.
Indeed, $s$ can be eliminated 
from (\ref{bohmj}) by writing
\begin{equation}
\frac{dx_a^{\mu}}{dx_b^{\nu}} = \frac{j_a^{\mu}}{j_b^{\nu}} .
\end{equation}
The trajectories are integral curves of the vector field 
${j_a^{\mu}}$. In this way, 
the Bohmian interpretation is relativistic covariant (since there is no 
preferred time coordinate), but nonlocal (since $Q$ is nonlocal)! 

The claim above that the Bohmian interpretation is covariant
requires an additional explanation.\footnote{I am 
grateful to D.~D\"urr and
S.~Goldstein for the discussions on that issue.}
Eq.~(\ref{bohmj}) can be viewed as an equation that determines 
{\em one} trajectory in the $4n$-dimensional configuration space. As an 
integral curve, such a trajectory is uniquely determined 
by one ``initial" position in the configuration space.
However, the physical trajectories are $n$ trajectories
in the 4-dimensional spacetime. The corresponding ``initial"
position corresponds to $n$ arbitrary points in spacetime 
that are proclaimed to have the same 
value of the parameter $s$. This arbitrariness corresponds 
to an arbitrary ``preferred" synchronization among $n$ particles.   
However, the theory is still 
covariant in the sense that there is no {\it a priori} 
preferred synchronization in the {\em equations of motion}. 
Instead, the structure of the covariant equations of motion 
is such that they allow a large number of different solutions.
Consequently, a choice of a ``preferred" synchronization
is only related to a choice of one of the {\em solutions}
to the equations of motion.\footnote{There is also a possibility 
for choosing a preferred foliation of spacetime in a 
{\em dynamical} way, as, for example, in \cite{durr99,nikolepjc}.} 
As known even from local classical theories,
the property of covariance refers to the equations of motion,
not to their solutions. 

A problem with such a covariant Bohmian interpretation 
is the following \cite{bern}.
In general, one does not know the initial probability 
distribution $\rho(x_1,\ldots ,x_n)$, so it seems that the theory 
does not have a predictive power. However,
in the next section, I demonstrate that, in some cases, there {\em are} 
measurable predictions, {\em not} equivalent to those of the conventional 
interpretation.

\section{One-particle case}

In this section, I discuss some physical consequences of the 
relativistic Bohmian equations of motion presented in the 
preceding section.
For simplicity, I consider the case of one particle.

Solutions of the Klein-Gordon equation are momentum eigenfunctions
\begin{equation}
\psi_{p}(x)=e^{-ip_{\mu}x^{\mu}}=e^{-i(\omega t-{\bf p}{\bf x})} ,
\end{equation}
where
\begin{equation}
\omega=\pm\sqrt{{\bf p}^2+m^2} .
\end{equation}
The wave function 
\begin{equation}
\psi(x)=\langle 0|\hat{\phi}(x)|1\rangle
\end{equation}
contains {\em only positive} frequencies. Nevertheless, 
if the superposition contains {\em different} positive frequencies,
$j^0$ may be negative at some regions of spacetime.
The Bohmian equation of motion can be written as 
\begin{equation}
\frac{dx^{\mu}}{ds} \propto j^{\mu} ,
\end{equation}
where the scalar of proportionality is an {\em arbitrary} 
scalar function of $x$. As the trajectory in spacetime is an 
integral curve of the vector field $j^{\mu}(x)$, this trajectory 
does not depend on the choice of that scalar function. 

The trajectories have several 
unusual properties. First, 
a particle moves {\em backwards in time} where $j^0<0$. 
Consequently, at a single time, particle may have two or more positions. 
In addition, at some points, particles move {\em superluminally}
(i.e., faster than $c\equiv 1$).
Nevertheless, this superluminal motion is consistent with 
relativity! The simplest way to understand this is to observe 
that the quantum Hamilton-Jacobi equation can be seen as an
equation of the form $p^{\mu}p_{\mu}=m^2_{\rm eff}$, where 
\begin{equation}
m^2_{\rm eff}(x)=m^2+
\frac{\partial^{\mu}\partial_{\mu} R(x)}{R(x)} 
\end{equation}
is the effective squared mass. It is clear that the effective
squared mass may be negative at some $x$, which corresponds to tachyons
at these $x$.

It is important to emphasize that the unusual properties above 
{\em are not in contradiction with observations, 
because the theory of quantum mesurements implies that the {\bf observed} 
velocities are never superluminal and the {\bf observed} position is 
always a single position}. 
For example, consider a measurement of velocity. The eigenfunction 
$\psi_{p}(x)$ is both the 4-momentum eigenstate and the
4-velocity eigenstate. Consequently, during a measurement we have 
an entangled state of the form
\begin{equation}
\Psi(x,y)=\int d^3p\: c_p\psi_p(x)\chi_p(y) .
\end{equation}
This implies that the {\em measured} velocity is always the velocity 
corresponding to one of the {\em eigenfunctions} $\psi_p(x)$, 
which is never superluminal. 
The case of measurement of the position is similar. As in the 
nonrelativistic case, the functions $\chi_a(y)$ form nonoverlapping
channels, so that a 
particle can enter {\em only one} of the localized channels. Consequently, 
when the position is measured, the particle cannot be found at two 
or more positions (channels) at the same time. 

We also emphasize that
superluminal velocities are consistent with causality.
That is, there are no causal paradoxes, because
the trajectories in spacetime are uniquely and self-consistently 
determined by the initial condition (see also \cite{nikolcaus}). 

Now we are ready to discuss the nontrivial measurable predictions of 
the relativistic Bohmian interpretation. 
Assume that initially
$j_0({\bf x},t_0)\geq 0$. In this case,
the initial probability density at $t_0$ is known;
\begin{equation}
\rho({\bf x},t_0)=j_0({\bf x},t_0) .
\end{equation}
Now assume that for $t>t_0$, $j_0<0$ at some regions. 
By explicitly {\em calculating the trajectories} for all possible 
initial coniditions at $t_0$, one can calculate 
$\rho({\bf x},t)$ at {\em any} $t$.  
We are interested in the {\em measurable} $\rho$, so we need to use 
the theory of quantum mesurements. 
Assume that the interaction with the apparatus that 
measures particle positions 
starts at some particular time $t_1$. 
In this case, the particles cannot move backwards in time 
for $t\geq t_1$. Consequently, some of the trajectories 
in the region $t_0<t<t_1$ cannot realize as physical trajectories
\cite{nikolfpl3}. As a result, for the measurable probability 
density at $t_1$ one finds \cite{nikolfpl3} 
\begin{equation}\label{rhomeas}
\rho({\bf x},t_1)=\left\{  
\begin{array}{ll}
j_0({\bf x},t_1) & \mbox{on $\Sigma'$}, \\
0 & \mbox{on $\Sigma^+\cup\Sigma^-$},
\end{array}   
\right.
\end{equation}
where $\Sigma^{\pm}$ and $\Sigma'$ are defined as follows.
$\Sigma^-$ is the set of all space points at $t_1$ at which 
$j_0<0$. $\Sigma^+$ is a similar set with $j_0>0$, such that
any point on $\Sigma^+$ is connected to a point on $\Sigma^-$   
via a trajectory in the region $t_0<t<t_1$. $\Sigma'$ is the 
set of all other points with $j_0\geq 0$ at $t_1$ that are 
not contained in $\Sigma^+$. (See also \cite{nikolfpl3} for a 
pictorial represesentation of this.) 
Eq.~(\ref{rhomeas}) says that,
at $t_1$, the particle {\em cannot} be found at 
any point in the region at which $j_0<0$, as well as at 
those points with $j_0>0$ that are connected 
with the $j_0<0$ region by a trajectory. 
This result cannot be obtained without 
calculating the trajectories. 
Possible experimental verification of the prediction (\ref{rhomeas})
would be 
the experimental proof that quantum particles {\em really} do have 
Bohmian trajectories!

\section{Conclusion and outlook}
  
From our results, two important conclusions can be drawn.
First, the
Bohmian interpretation can be formulated such that 
it is both relativistic covariant and nonlocal. 
Second, for the case $j_0<0$, the
conventional interpretation of relativistic quantum mechanics  
does not have {\em any} prediction on the distribution of particle positions. 
On the other hand, the Bohmian interpretation {\em does}! 

Our results also suggest some
prospects for the future experiments. I believe that
we should start thinking how to prepare states with $j_0<0$
and how to measure the particle distribution for this case.
Since there are no experimentally 
accesible elementary particles with spin zero, perhaps the 
experiments could be performed with photons, which also 
satisfy a second-order differential equation similar to the 
Klein-Gordon equation, so that negative particle densities 
also take place.
Even if the result will not confirm the Bohmian prediction 
considered in this work, the point is that {\em there is no
standard prediction} on that issue. Consequently,
the result will certainly tell us something new and fundamental about 
relativistic quantum mechanics.

\section*{Acknowledgments}

This work was supported by the Ministry of Science and Technology of the
Republic of Croatia under Contract No.~0098002.


\begin{thebibliography}{99}

\bibitem{bohm}
D.~Bohm, Phys.~Rev.~{\bf 85}, 166, 180 (1952).
\bibitem{bohmPR1}
D.~Bohm and B.~J.~Hiley,
{\it Phys.~Rep.}~{\bf 144}, 323 (1987).
\bibitem{bohmPR2}
D.~Bohm, B.~J.~Hiley, and P.~N.~Kaloyerou,
{\it Phys.~Rep.}~{\bf 144}, 349 (1987).
\bibitem{holPR}
P.~R.~Holland, {\it Phys.~Rep.}~{\bf 224}, 95 (1993).
\bibitem{holbook}
P.~R.~Holland, {\it The Quantum Theory of Motion}
(Cambridge University Press, Cambridge, 1993).
\bibitem{nikolfpl1}
H.~Nikoli\'c, {\it Found. Phys. Lett.} {\bf 17}, 363 (2004).
\bibitem{nikolfpl3}
H.~Nikoli\'c, {\it Found. Phys. Lett.} {\bf 18}, 549 (2005).
\bibitem{val} 
A.~Valentini, {\it Phys.~Lett.}~{\bf A} 156, 5 (1991).
\bibitem{durr1}
D.~D\"urr, S.~Goldstein, and N.~Zangh\`i,
{\it J.~Stat.~Phys.}~{\bf 67}, 843 (1992).
\bibitem{durr2}
D.~D\"urr, S.~Goldstein, and N.~Zangh\`i,
{\it Phys.~Lett.}~A {\bf 172}, 6 (1992).
\bibitem{nikolfpl2}
H.~Nikoli\'c, {\it Found. Phys. Lett.} {\bf 18}, 123 (2005).
\bibitem{nikolplb}
H.~Nikoli\'c, {\it Phys. Lett. B} {\bf 527}, 119 (2002).
\bibitem{nikolijmpd}
H.~Nikoli\'c, {\it Int. J. Mod. Phys. D} {\bf 12}, 407 (2003).
\bibitem{nikolgrg}
H.~Nikoli\'c, {\it Gen. Rel. Grav.} {\bf 37}, 297 (2005).
\bibitem{durr99}
D.~D\"urr, S.~Goldstein, K.~M\"unch-Berndl, and N.~Zangh\`i,
{\it Phys.~Rev.}~A {\bf 60}, 2729 (1999).
\bibitem{nikolepjc}
H.~Nikoli\'c, {\it Eur. Phys. J. C} {\bf 42}, 365 (2005).
\bibitem{bern}
K.~Berndl, D.~D\"urr, S.~Goldstein, and N.~Zangh\`i,
{\it Phys.~Rev.}~A {\bf 53}, 2062 (1996).
\bibitem{nikolcaus}
H.~Nikoli\'c, gr-qc/0403121.

\end{thebibliography}
\end{document}